\documentclass[aps,prd,superscriptaddress,twocolumn,floatfix,nofootinbib,preprintnumbers,showpacs]{revtex4}
\usepackage{amsmath,multirow}
\usepackage{graphicx,tabularx}
\usepackage{url}

\newcommand{\etal}{\textit{et al. }}
\newcommand{\ie}{{\sl i.e. }}
\newcommand{\eg}{{\sl e.g. }}
\newcommand{\mbb}{m_{bb}^\text{rec}}
\newcommand{\mt}{m_t^\text{rec}}
\newcommand{\mw}{m_W^\text{rec}}

\newcommand{\gev}{{\ensuremath\rm GeV}}

\newcommand{\fb}{{\ensuremath\rm fb}}

\newcommand{\ifb}{{\ensuremath\rm fb^{-1}}}

\begin{document}

\preprint{KA-TP-12-2009}



\title{Fat Jets for a Light Higgs}

\author{Tilman Plehn}
\affiliation{Institut f\"ur Theoretische Physik, Universit\"at Heidelberg, Germany}

\author{Gavin P. Salam}
\affiliation{LPTHE, UPMC Univ Paris 6, CNRS UMR 7598, Paris, France}

\author{Michael Spannowsky}
\affiliation{Institut f\"ur Theoretische Physik, Universit\"at Karlsruhe, KIT, Germany and \\
             Department of Physics and Institute of Theoretical Science, University of Oregon, Eugene, USA}

\begin{abstract}
  At the LHC associated top quark and Higgs boson production with a
  Higgs decay to bottom quarks has long been a heavily disputed search
  channel. Recently, it has been found to not be viable. We show how
  it can be observed by tagging massive Higgs and top jets. For this
  purpose we construct boosted top and Higgs taggers for Standard Model
  processes in a complex QCD environment.
\end{abstract}

\maketitle


The main task of the LHC is to understand electroweak symmetry
breaking, \eg by confirming or modifying the minimal Higgs mechanism
of the Standard Model~\cite{higgs,review}. In the Standard Model as
well as its typical perturbative extensions, electroweak precision
data clearly prefer a light Higgs boson~\cite{lepewwg}, most likely
well below the threshold of Higgs decays to $W$ bosons. If only
because 68~\% of light Higgs bosons ($m_H = 120$~GeV) decay to bottom
quarks~\cite{hdecay}, we should look for this Higgs signature.\bigskip

Over the past years, Higgs search strategies based on different
production mechanisms have been developed.

The dominant gluon-fusion production process cannot be combined with a
decay to bottom quarks, because of its overwhelming QCD
background. For this production process all hopes rest on the Higgs
decay to photons~\cite{gf_gamma} with its challenging
signal-to-background ratio.

Higgs production in weak boson fusion with a decay to bottoms
challenges the Atlas and CMS triggers~\cite{wbf_b}. Combined with
a decay to taus instead, it is one of the discovery
channels~\cite{wbf_tau} --- provided analysis techniques like a
central jet veto and collinear $\tau \tau$ mass reconstruction work in
the QCD environment of the LHC.

While at the Tevatron the associated $ZH$ and $WH$ production serves
as a discovery channel, at the LHC it is plagued by QCD
backgrounds. Nevertheless, a recent study has shown that using a fat
Higgs jet --- \ie a jet from a massive particle decay with subjet
structure --- we can extract $WH/ZH$ production with $H \to b \bar{b}$
for a Higgs mass of 120~GeV at the $\sim 4\sigma$ level using $30~\ifb$
of data~\cite{gavin,giacinto}.

Additional search channels like weak-boson-fusion production of
$\gamma H$~\cite{wbf_gammah} or $WH$~\cite{wbf_wh} final states
combined with a decay $H \to b \bar{b}$ might be visible, but lack a
final experimental word. It is clear, though, that none of them will
lead to a discovery in the first years of LHC running.\bigskip

Last but not least, the associated production of a top quark with a
Higgs boson at the LHC has a long history, usually in combination with
a Higgs decay to bottoms. At some point it was expected to be the
leading discovery channel for a light Higgs boson~\cite{tth_working},
but recently it has been removed from the Higgs discovery plots by
Atlas and CMS~\cite{tth_dead}. Without systematic uncertainties, Atlas
quotes a significance of 1.8 to $2.2\sigma$ for $30~\ifb$. Due to a
(too) low signal-to-background ratio $S/B \sim 1/9$ this channel might
not reach a $5\sigma$ significance for any luminosity. The main
problems are the combinatorial background of bottom jets and 
the lack of a truly distinctive kinematic feature of the Higgs decay
jets. \bigskip

Any meaningful analysis of the Higgs sector has to test the Yukawa
nature of the Higgs--fermion couplings. In addition to the bottom
Yukawa coupling discussed above we expect to extract the top Yukawa
coupling from one-loop contribution to the higher-dimensional $ggH$
and $\gamma \gamma H$ couplings. However, any kind of new heavy
particle will also contribute to both of them, which makes it hard to
perform a model independent top coupling measurement. A measurement
based on a direct (\ie tree level) production process is the only way
to reliably measure the top Yukawa. All of these arguments point to
\begin{equation}
pp \to t \bar{t} H \to t \bar{t} \; b \bar{b}
\end{equation}
as a prime ingredient for understanding the Higgs sector at the
LHC~\cite{sfitter}.\bigskip

In this paper we show how, using fat jets, this Standard Model search
channel can indeed be extracted with reasonable statistical
significance and most importantly a much reduced sensitivity on
systematics. The combinatorial problem in the signal we solve by the
construction of two fat jets; based on those we find plenty of kinematic
distributions which separate signal and background.  

Fat jets have been studied in the framework of searches for strongly
interacting $W$ bosons~\cite{fatjet_w}, supersymmetric
particles~\cite{fatjet_susy}, heavy resonances decaying to strongly
boosted top quarks~\cite{tt_resonance}, as well as the $WH/ZH$ search
mentioned above~\cite{gavin,giacinto}. For leptonic top quarks they
are similar to complex mass and momentum reconstruction
tools~\cite{top_rec_semilep}.  Top
taggers~\cite{david_e,top_tag} have been studied  in high-$p_T$
contexts, 
but differ in their applicability once the top quarks
are only slightly boosted, $E/m_t \gtrsim 1$.
Therefore, we construct
Standard-Model Higgs and top taggers for tagging in busy environments
at moderately high 
$p_T$ and show how fat Higgs as
well as top jets can be used to identify a Standard Model Higgs
signature.\bigskip

\underline{Signal and backgrounds} --- We consider associated top and
Higgs production with one hadronic 
and one
leptonic 
top decay. The latter allows the events to pass the Atlas and CMS
triggers. The main backgrounds are
\begin{alignat}{5}
pp &\to t \bar{t} b \bar{b}
&&\text{irreducible QCD background}
\notag \\
pp &\to t \bar{t} Z
&&\text{irreducible $Z$-peak background}
\notag \\
pp &\to t \bar{t} +\text{jets} \qquad
&& \text{include fake bottoms}
\label{eq:back}
\end{alignat}
To account for higher-order effects we normalize our total signal rate
to the next-to-leading order prediction of 702~fb for
$m_H=120$~GeV~\cite{signal_nlo}.
The $t \bar{t} b \bar{b}$ continuum background we normalize to 2.6~pb
after the acceptance cuts $|y_b|<2.5, p_{T,b}>20$~GeV and
$R_{bb} > 0.8$ of Ref.~\cite{ttbb_nlo}.
This conservative rate estimate for very hard events implies a $K$
factor of $\sigma_\text{NLO}/\sigma_\text{LO} = 2.3$ which we need to
attach to our leading-order background simulation --- compared to
$K=1.57$ for the signal.  Finally, the $t \bar{t} Z$ background at NLO
is normalized to 1.1~pb~\cite{ttz_nlo}.
For $t \bar{t}$ plus jets production we do not apply a higher-order
correction because the background rejection cuts drives it into
kinematic configuration in which a constant $K$ factor cannot be used.
Throughout this analysis we use an on-shell top mass of 172.3~GeV.
All hard processes we generate using MadEvent~\cite{madevent}, shower
and hadronize via Herwig++~\cite{herwig} (without $g \to b \bar{b}$
splitting) and analyze with FastJet~\cite{fastjet}. We have verified
that we obtain consistent results for signal and background using
Alpgen~\cite{alpgen} and Herwig~6.5~\cite{f_herwig}\bigskip

An additional background is $W$+jets production. The $Wjj$ rate starts
from roughly 15~nb with $p_{T,j}>20$~GeV. Asking for two very hard
jets, mimicking the boosted Higgs and top jets, and a leptonic $W$
decay reduces this rate by roughly three orders of magnitude. Our top
tagger described below gives a mis-tagging probability around $5\%$
including underlying event, the Higgs mass window another reduction by
a factor $1/10$, \ie the final $Wjj$ rate without flavor tags ranges
around 100~fb.

Adding two bottom tags we expect a purely fake-bottom contribution
around 0.01~fb.  To test the general reliability of bottom tags in QCD
background rejection we also simulate the $Wjj$ background including
bottom quarks from the parton shower and find a remaining background
of $\mathcal{O}(0.1~\fb)$, well below $10\%$ of the $t\bar{t}$+jets
background already for two bottom tags. For three bottom tags it is
essentially zero, so we neglect it in the following.

The charm-flavored $Wcj$ rate starts off with 1/6 of the purely
mis-tagged $Wjj$ rate. A tenfold mis-tagging probability still leaves
this background well below the effect of bottoms from the parton
shower.  Finally, a lower limit $\mbb > 110$~GeV keeps us safely away
from CKM-suppressed $W \to b \bar{c}$ decays where the charm is
mis-identified as a bottom jet.\bigskip

\begin{figure}[t]
\includegraphics[width=0.40\textwidth]{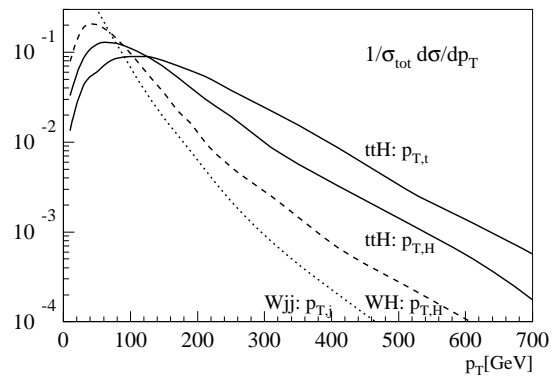}
\vspace*{-4mm}
\caption{Normalized top and Higgs transverse momentum spectra in $t
  \bar{t} H$ production (solid). We also show $p_{T,H}$ in $W^- H$
  production (dashed) and the $p_T$ of the harder jet in $W^- jj$
  production with $p_{T,j} > 20$~GeV (dotted).}
\label{fig:pt}
\end{figure}

\underline{Search strategy} --- The motivation for a $t \bar{t} H$
search with boosted heavy states can be seen in Fig.~\ref{fig:pt}: the
leading top quark and the Higgs boson both carry sizeable transverse
momentum. We therefore first cluster the event with the Cambridge/Aachen (C/A) jet
algorithm~\cite{ca_algo} using $R=1.5$ and require two or more hard jets
and a lepton satisfying:
\begin{alignat}{5}
p_{T,j} &> 200~\gev \qquad \qquad
&|y^{(H)}_j| &< 2.5 \qquad |y^{(t)}_j| < 4
\notag \\p_{T,\ell} &> 15~\gev 
&|y_\ell| &< 2.5 \; .
\label{eq:acc}
\end{alignat}
The maximum Higgs jet rapidity $y_J^{(H)}$ is limited by the
requirement that it be possible to tag its $b$-content.  For lepton
identification and isolation we assume an 80\% efficiency, in
agreement with what we expect from a fast Atlas detector simulation.  The outline of
our analysis is then as follows (cross sections at various stages are
summarized in Tab.~\ref{tab:eff}):\bigskip

\noindent (1) one of the two jets should pass the top tagger
(described below).  If two jets pass we choose the one whose top
candidate is closer to the top mass.

\noindent (2) the Higgs tagger (also described below) runs over all
remaining jets with $|y| < 2.5$. It includes a double bottom tag.

\noindent (2') a third $b$ tag can be applied in a separate jet analysis
after removing the constituents associated with the top and Higgs.

\noindent (3) to compute the statistical significance we require $\mbb
= m_H \pm 10$~GeV.\bigskip

In this analysis, QCD $t \bar{t}$ plus jets production can fake the
signal assuming three distinct topologies: first, the Higgs candidate jet
can arise from two mis-tagged QCD jets. The total rate without
flavored jets exceeds $t \bar{t} b \bar{b}$ production by a factor of
200. This ratio can be balanced by the two $b$ tags inside the Higgs
resonance. Secondly, there is an $\mathcal{O}(10\%)$ probability for
the bottom from the leptonic top decay to leak into the Higgs jet and
combine with a QCD jet,
to fake a Higgs candidate. This topology is the most dangerous
and can be essentially removed by a third $b$ tag outside the Higgs
and top substructures. Finally,  the bottom from the hadronic top
can also leak into the Higgs jet after being replaced by a QCD jet with the
appropriate kinematics in the top reconstruction.\bigskip

These three distinct topologies appear in the $t\bar{t}$ background
because of the unusually large QCD jet activity which we corresponds
to the huge QCD correction to the total rate.  The impact of these
background configurations on our analysis critically depends on the
detailed simulation of QCD jet radiation in $t \bar{t}$ events.  We
therefore perform our entire analysis for the minimal two $b$ tags as
well as for a safe scenario with three $b$ tags, to achieve a maximal
reduction of this background.\bigskip

\begin{table}[t]
\begin{tabular}{l|r|rrr}
 & signal & $t \bar{t} Z$ & $t \bar{t} b \bar{b}$ & $t \bar{t}$+jets \\
\hline
events after acceptance eq.(\ref{eq:acc})   & 24.1  &  6.9  & 191    & 4160 \\
events with one top tag                     & 10.2  &  2.9  &  70.4  & 1457 \\
events with $\mbb = 110 - 130$~GeV          &  2.9  & 0.44  &   12.6 & 116 \\
corresponding to subjet pairings            &  3.2  & 0.47  &   13.8 & 121 \\ 
\hline
subjet pairings two subjet $b$ tags         & 1.0  & 0.08 &  2.3 &  1.4 \\
including a third $b$ tag                    & 0.48   & 0.03 & 1.09   & 0.06  \\            
\end{tabular}  
\caption{Number of events or $\mbb$ histogram entries per $1~\ifb$
  including underlying event, assuming $m_H=120$~GeV. The third row
  gives the number of events with at least one subjet pairing in the
  Higgs mass window while the fourth row (and below) gives the number
  of entries according to our algorithm based on the three leading
  modified Jade distances.}
\label{tab:eff}
\end{table}

\underline{Top and Higgs taggers} --- In contrast to other Higgs
physics~\cite{gavin} or new physics~\cite{fatjet_w,fatjet_susy}
applications our Higgs and top taggers cannot rely on a clean QCD
environment: on the one hand their initial cone size has to be large
enough to accommodate only mildly boosted top and Higgs states, so
additional QCD jets will contaminate our fat jets~\cite{pruning}. On
the other hand, the small number of signal events does not allow any
sharp rejection cuts for dirty QCD events. Therefore, the taggers need
to be built to survive busy LHC events.\bigskip

Our starting point is the C/A jet algorithm with $R=1.5$. For a top
candidate, which typically has a jet mass above 200~GeV, we assume that
there could be a complex hard substructure inside the fat jet. 
To reduce this fat jet to the relevant substructures we apply the
following recursive procedure.  The last clustering of the jet $j$ is
undone, giving two subjets $j_1,j_2$, ordered such that $m_{j_1} >
m_{j_2}$. If $m_{j_1}> 0.8~m_j$ (\ie $j_2$ comes from the underlying
event or soft QCD emission) we discard $j_2$ and keep $j_1$, otherwise
both $j_1$ and $j_2$ are kept; for each subjet $j_i$ that is kept, we
either add it to the list of relevant substructures (if $m_{j_i} <
30~\gev$) or further decompose it recursively.

In the resulting set of relevant substructures, we examine all
two-subjet configurations to see if they could correspond to a $W$
boson: after filtering as in Ref.\cite{gavin} to reduce contamination from the
underlying event, the mass of the substructure pair should be in the
range $\mw = 65 - 95$~GeV (shown in Fig.~\ref{fig:topmass}).  To tag
the top quark, we then add a third subjet and, again after
filtering~\cite{gavin}, require $\mt = 150 - 200$~GeV.  We
additionally require that the $W$ helicity angle $\theta$ with respect
to the top candidate satisfies $\cos \theta <0.7$, as in
Ref.\cite{david_e}.  For more than one top tag in the event we choose
the one with the smaller $|\mt -m_t^\text{pole}| + |\mw
-m_W^\text{pole}|$.  The resulting top tagging efficiency in the
signal, including underlying event, is $43\%$, with a 5\%
mis-tagging probability in $W$+jets events. Note that these values
hold for only slightly boosted tops and in a particularly complex QCD
environment.\bigskip

In contrast to the top tagger which identifies a top quark using its
known mass and properties, our Higgs tagger~\cite{gavin} has to search
for a Higgs peak in the reconstructed $\mbb$ without any knowledge of
the Higgs mass. We use the same decomposition procedure described
above (but now with a mass cutoff at 40~GeV and a mass drop threshold
of 0.9). We then order all possible
pairs of subjets by the modified Jade distance~\cite{fatjet_susy}
\begin{alignat}{5}
J = p_{T,1} p_{T,2} \left( \Delta R_{12} \right)^4 \; ,
\end{alignat}
similar to the mass of the
hard splitting, but shifted
towards larger jet separation. The three leading pairings we filter
and keep for the Higgs mass reconstruction. For these events we
explicitly confirm that indeed we are dominated by $p_{T,H} \gtrsim
200$~GeV.\bigskip

\begin{figure}[t]
\includegraphics[width=0.40\textwidth]{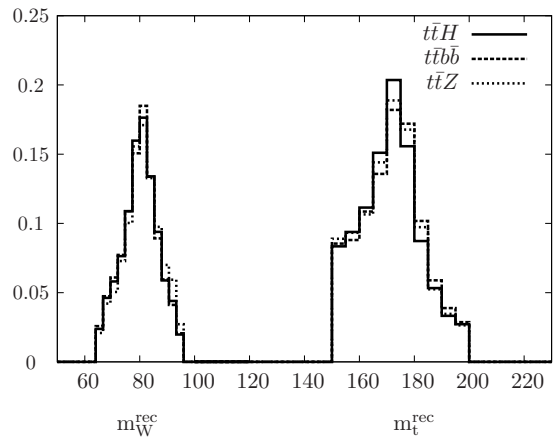}
\vspace*{-4mm}
\caption{Individually normalized $\mw$ and $\mt$ distributions for
  signal and background (with underlying event).}
\label{fig:topmass}
\end{figure}

\underline{Double vs triple bottom tag} --- At this stage we have not yet
included any flavor tags to control the $t\bar{t}$+jets and $W$+jets
backgrounds. To reduce the leading $t\bar{t}jj$ topology we first
require two bottom tags for the substructure pairings reconstructing
the Higgs.  Based on the detector-level study~\cite{giacinto} we
assume a 70\% efficiency with a 1\% mis-tagging probability for $b$
tags of filtered Higgs subjets.

We then apply a $\pm 10$~GeV mass window, after checking that the
tails of the signal distribution drop sharply in particular towards
larger mass values.  In the double $b$-tag analysis we find for an
integrated luminosity of $100~\ifb$:\\[-9mm]

\begin{center}
\begin{tabular}{r|rrr|r}
 & $S$ & $B$ & $S/B$ & $S/\sqrt{B}$ \\
\hline
$m_H = 115$~GeV & 120  & 380 & 1/3.2 & 6.2 \\
      $120$~GeV & 100  & 380 & 1/3.8 & 5.1 \\
      $130$~GeV & 51   & 330 & 1/6.5 & 2.8
\end{tabular}
\end{center}

This result shows that we can extract the $t\bar{t}H$ signal with high
significance. On the other hand, similar to the original Atlas and CMS
analyses it suffers from low $S/B$, the impact of the poorly
understood $t\bar{t}$+jets background with its different kinematic
topologies, its large theory uncertainty and potentially large
next-to-leading order corrections, and the missing underlying
event.\bigskip

To improve the signal-to-background ratio $S/B$ and remove the impact
of the $t \bar{t}$+jets background (at the expense of the final
significance) we can apply a third $b$ tag.  Targeting the second
$t\bar{t}$+jets topology we remove the Higgs and top constituents from
the event and cluster the remaining particles into jets using the C/A
algorithm with $R=0.6$, considering all jets with
$p_T>30$~GeV. Amongst these jets we require one $b$ tag with $\eta <
2.5$ and a distance $\Delta R_{b,j}>0.4$ to the Higgs and top
subjets, assuming 60\% efficiency and 2\% purity. The last row of
Table~\ref{tab:eff} confirms that requiring three bottom tags leaves
the continuum $t \bar{t} b \bar{b}$ production as the only relevant
background.\bigskip

In Fig.~\ref{fig:higgsmass} we show the signal from the three leading
(by modified Jade distance) $\mbb$ entries of double-$b$-tagged
combinations; our Higgs tagger returns a sharp mass peak.  The bigger
tail towards small $\mbb$ we can reduce by only including the two
leading jet combinations. This does not change the significance but
sculpts the background more.  Assuming that at this stage we will know
the Higgs mass, we estimate the background from a clean right and a
reasonably clean left side bin combined with a next-to-leading order
prediction. The result of the triple $b$-tag analysis is then (again
assuming $100~\ifb$):\\[-9mm]

\begin{center}
\begin{tabular}{r|rrr|r}
 & $S$ & $B$ & $S/B$ & $S/\sqrt{B}$ \\
\hline
$m_H = 115$~GeV & 57 & 118 & 1/2.1 & 5.2 (5.7) \\
      $120$~GeV & 48 & 115 & 1/2.4 & 4.5 (5.1) \\
      $130$~GeV & 29 & 103 & 1/3.6 & 2.9 (3.0)
\end{tabular} 
\end{center}

The numbers in parentheses are without underlying event. While
removing the highly uncertain $t \bar{t}$+jets background has indeed
lowered the final significance, the background of the three $b$-tag
analysis is completely dominated by the well-behaved $t \bar{t} b
\bar{b}$ continuum production.\bigskip

\begin{figure}[t]
\includegraphics[width=0.39\textwidth]{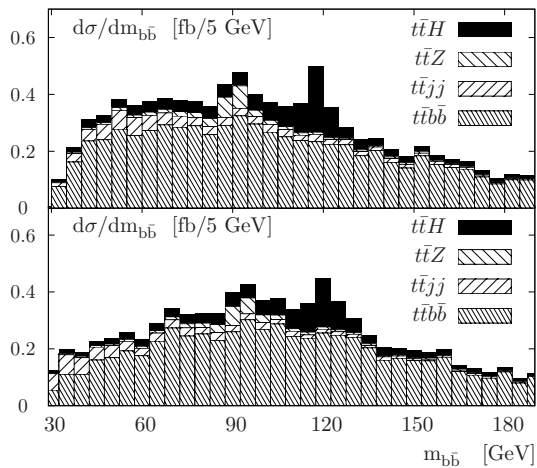}
\vspace*{-3mm}
\caption{Reconstructed bottom-pair mass $\mbb$ for signal ($m_H =
  120$~GeV) and backgrounds without (upper) and including (lower)
  underlying event. The distributions shown include three $b$ tags.}
\label{fig:higgsmass}
\end{figure}

\underline{Further improvements} --- One of the problems in this
analysis is that higher-order QCD effects harm its reach.  Turning
this argument around, we can use the additional QCD activity in the
signal and continuum $t \bar{t} b \bar{b}$ background to improve our
search. Before starting with the fat-jet analysis we can for example
analyze the four leading jets with a radius $R=0.6$ and $p_T<40$~GeV
and require a set of jet-jet and jet-lepton separation
criteria~\cite{single_top}: we reject any event for which one of the three
conditions holds
\begin{alignat}{7}
\cos \theta^*_{j_2 j_1} &< -0.4 \qquad && \text{and} \qquad 
&& \Delta {k_T}_{j_3 \ell} \; \epsilon \; [70,160]~\gev
\notag \\
\cos \theta^*_{j_3 \ell} &> 0.4 && \text{and} 
&&\Delta R_{j_2 j_3} > 2.5
\notag \\
\Delta R_{j \ell} &> 3.5 &&&& \hspace*{-2cm} \text{for any of the four leading jets.} 
\end{alignat}
$\theta^*_{P_1 P_2}$ is the angle between $\vec{p}_1$ in the
center-of-mass frame of $P_1 + P_2$ and the center of mass direction
$(\vec{p}_1+\vec{p}_2)$ in the lab frame. It is not symmetric in its
arguments; if the two particles are back to back and $|\vec{p}_1| >
|\vec{p}_2|$ it approaches $\cos \theta^* =1$, whereas for
$|\vec{p}_1| < |\vec{p}_2|$ it becomes $-1$~\cite{single_top}. The
$k_T$ distance between two particles is $(\Delta {k_T}_{j \ell})^2 = \min( p_{T,j}^2, p_{T,\ell}^2)
\Delta R_{j \ell}^2$.  At this stage and with our limited means of detector
simulation this QCD pre-selection at least shows that there are
handles to further improve $S/B$ from 1/2.4 to roughly 1/2 (for $m_H =
120$~GeV) with hardly any change to the final significance.\bigskip

In addition, we can envisage improving the analysis in several ways 
in the context of a full experimental study, including data to help
constrain the simulations:

\noindent (1) Replace the $\mbb$ side bins by a likelihood analysis of
the well-defined alternative of either $t\bar{t}H$ signal or
$t\bar{t}b\bar{b}$ continuum background after three $b$ tags. This
increases the final number of events, our most severe limitation.

\noindent (2) Provided the events can be triggered/tagged, include two
hadronic or two leptonic top decays. This more than triples the
available rate and includes a combinatorical advantage of requiring
one of two tops to be boosted.

\noindent (3) Without cutting on missing energy as part of the
acceptance cuts use its measurement within errors to assign the
correct jet to the leptonic top and become less dependent on the third
$b$ tag.\bigskip

\underline{Outlook} --- In this paper we have presented a new strategy
to extract the Higgs production process $t\bar{t}H$ with the decay $H
\to b \bar{b}$ at the LHC. After long debates this signature has
recently been abandoned by both LHC experiments, even though it would
be an especially useful ingredient to a complete Higgs sector analysis
at the LHC~\cite{sfitter}.  We propose two analysis strategies based
on a boosted Higgs boson~\cite{gavin} and a boosted top quark; one
with a double and one with a triple $b$ tag. The latter compensates
its reduced statistical significance with a strongly reduced
dependence on systematic uncertainties. The only remaining background
after three $b$ tags is continuum $t\bar{t}b\bar{b}$ production with
accessible side bins.\bigskip

For an integrated luminosity of $100~\ifb$ and a Higgs mass of 120~GeV
our three $b$-tag analysis gives a statistical significance of at
least $4.5\sigma$ and a signal-to-background ratio of at least
$S/B=1/2.4$.  The signal-to-background ratio can be further improved
using the structure of the QCD radiation for signal and
background. Combinatorial backgrounds are not a problem, and we find a
multitude of distributions distinguishing between signal and continuum
background.\bigskip

\underline{Acknowledgments} --- TP would like to thank the University
of of Washington and their Higgs workshop, supported in part by the
DOE under task TeV of DE-FGO3-96-ER40956, for many discussions. MS and
GPS would like to thank the LHC New Physics Forum in Heidelberg, where
the idea was born. Last but not least, we would like to thank Jon
Butterworth, Giacinto Piacquadio, Brock Tweedie, Gregory Soyez, Matt
Strassler, Steve Ellis, Stefan Gieseke, Christoph Hackstein, and
Ulrich Uwer for help and enjoyable discussions. GPS would like to
thank the French ANR for support under contract ANR-09-BLAN-0060.



\end{document}